\documentclass[12pt,twoside]{article}
\usepackage{fleqn,espcrc1}
\usepackage{graphicx}

\newcommand{\be}{\begin{eqnarray}}
\newcommand{\ee}{\end{eqnarray}}

\newcommand{\AmS}{{\protect\the\textfont2
  A\kern-.1667em\lower.5ex\hbox{M}\kern-.125emS}}
\title{Drell-Yan and $J/\psi$
production in high energy
proton-nucleus and nucleus-nucleus collisions\footnote{Presented by
J. Kapusta and C. Gale.}}

\author{Charles Gale\address{Physics Department, McGill University\\
Montreal, Quebec H3A 2T8, Canada}, Sangyong Jeon\address{Nuclear
Science Division,  Lawrence Berkeley National Laboratory\\
Berkeley, CA 94720, USA}, and Joseph Kapusta\address{School of Physics and
Astronomy,  University of Minnesota\\
Minneapolis, MN 55455, USA}}

\begin{document}

\maketitle
\begin{abstract}

The distributions of outgoing protons and charged hadrons in high
energy proton-nucleus collisions are described rather well by a
linear extrapolation from proton-proton collisions.  This linear extrapolation
is applied to precisely measured Drell-Yan cross sections for 800 GeV
protons incident on a variety of nuclear targets. The deviation from
linear scaling in the atomic number $A$ can be accounted for by
energy degradation of the proton as it passes through the nucleus if
account is taken of the time delay of particle production due to
quantum coherence.  We infer an average proper coherence time of
0.4$\pm$0.1 fm/c.  Then we apply the linear
extrapolation to measured $J/\psi$ production cross sections for
200 and 450 GeV/c protons incident on a variety of nuclear targets.
Our analysis takes into account energy loss of the beam proton, the time
delay of particle production due to quantum coherence, and absorption
of the $J/\psi$ on nucleons.  The best representation is obtained for a
coherence time of 0.5 fm/c, which is consistent with Drell-Yan production,
and an absorption cross section of 3.6 mb, which is consistent with the
value deduced from photoproduction of the $J/\psi$ on nuclear targets.
Finally, we compare to recent $J/\psi$ data from
S+U and Pb+Pb collisions at the SPS.  The former are reproduced
reasonably well with no new parameters, but not the latter.

\end{abstract}

\section{INTRODUCTION}

There are two extreme limits of a projectile scattering from a nucleus.
When the cross section of the projectile with a nucleon is very small,
as is the case for neutrinos, Glauber theory says that the cross section
with a nuclear target of atomic number $A$ grows linearly with $A$.  When
the cross section with an individual nucleon is very large, as is
the case for pions near the delta resonance peak, the nucleus appears
black and the cross section grows like $A^{2/3}$.  A more interesting
case is the production of lepton pairs with large invariant mass,
often referred to as Drell-Yan, in proton-nucleus
collisions.  Both the elastic and inelastic cross sections for
proton-nucleon scattering are relatively large, but the partial
cross section to produce a high mass lepton pair, being electromagnetic
in origin, is relatively small.  Experiments have shown that the
inclusive Drell-Yan cross section grows with $A$ to a power very
close to 1.  The theoretical interpretation is that the hard
particles, the high invariant mass lepton pairs, appear first
and the soft particles, the typical mesons, appear later
due to quantum-mechanical interference, essentially the uncertainty
principle.  These quantum coherence requirements also lead to the
Landau-Pomeranchuk-Migdal effect \cite{lpm}.  Deviations from the
power 1 by high precision Drell-Yan experiments \cite{E772}
at Fermi National Accelerator Laboratory (FNAL) suggest
that it may be possible to infer a finite numerical value for the
coherence time.  That is one of our goals.

Another of our goals is to take this coherence time effect into account when
extracting an absorption cross section for $J/\psi$ on nucleons for $J/\psi$
particles produced in high energy proton-nucleus collisions. After
extracting this absorption cross section, we use it to carry
out parameter-free calculations of $J/\psi$ production in S+U and Pb+Pb
collisions to help us ascertain the likelihood that new physics must be
invoked, that is, quark-gluon plasma.

Before addressing these goals it is imperative to have a basic description of
high energy proton-nucleus collisions which reproduces the essential data on
outgoing baryons and mesons.  We will use one particular theoretical approach,
referred to as LEXUS.  Using input from nucleon-nucleon collisions together
with geometry, LEXUS does describe the data very well.  See
\cite{lexus,prl,kkg}.  However, it important to realize that any model or
extrapolation which incorporates the same basic features will lead to the
same conclusions we find here.

\section{PROTON-NUCLEUS COLLISIONS}
\label{sect:pA}

\subsection{Drell-Yan production}

Consider a description of the Drell-Yan process in a proton-nucleus
collision.  One limit is full
energy degradation of the proton as it traverses the nucleus.  Produced
hadrons appear immediately with zero coherence time, causing the proton
to have less energy available to produce a Drell-Yan pair at the backside
of the nucleus.  The other limit is usually referred to as Glauber,
although this is a bit of a misnomer.  Produced hadrons, being soft
on the average, do not appear until after the hardest particles, the
Drell-Yan pair, have already appeared.  This is the limit of a very
large coherence time, and it allows the proton to produce the Drell-Yan
pair anywhere along its path with the full incident beam energy.
An intermediate case is one of finite, nonzero coherence time.
By the time the proton wants to make a Drell-Yan pair on the backside
of the nucleus, hadrons have already appeared from the first collision
but not from the second.  Therefore the proton has more energy available
to produce the Drell-Yan pair than full zero coherence time but less
energy than with infinite coherence time.  This ought to result
in an $A$ dependence less than 1, with the numerical value determined
by the coherence time.  It is instructive to contemplate the relative
importance of energy loss and coherence time for an 800 GeV proton
incident on a {\em very} large nucleus, such as a neutron star: Can
one imagine the proton reaching the backside of a neutron star and
producing a Drell-Yan pair without having suffered {\em any} energy loss?

Consistent with our philosophy to describe everything in terms of
hadronic variables we should use a parametrization of measured
Drell-Yan cross sections in pp and pn collisions.  However, we
need these over a very broad energy range because of the decreasing
energy of the proton as it cascades through the nucleus, and such
broad measurements have not been made.  Therefore, we compute the
Drell-Yan yields in individual pp and pn collisions using the parton
model with the GRV structure functions \cite{grv94} to leading order
with a K factor.  These structure functions distinguish between pp
and pn collisions.  We have compared the results to pp collisions at
the same beam energy of 800 GeV \cite{800pp} and
found the agreement to be excellent for all values of $x_F$.

\begin{figure}[htb!]
\centerline{\includegraphics[angle=90,width=4in]{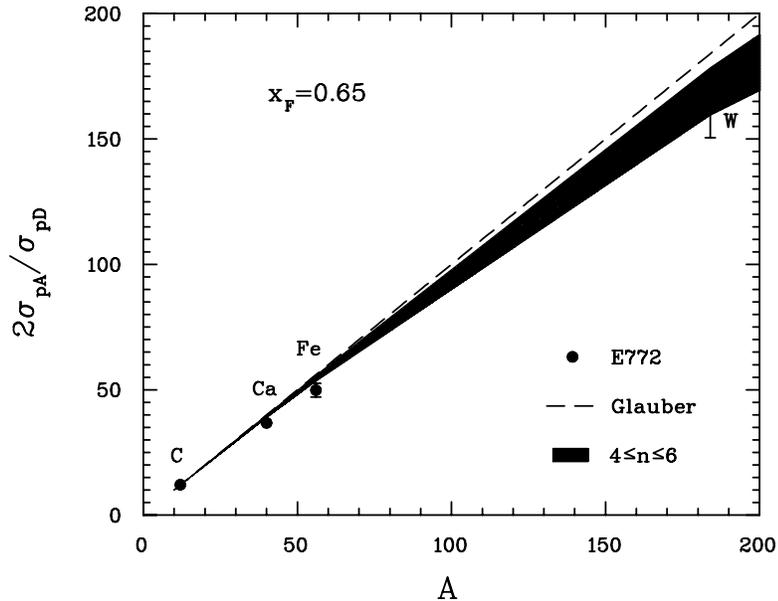}}
\caption{ The ratio of the pA Drell-Yan cross section to the
proton-deuterium cross section divided by 2 for a beam energy
of 800 GeV.  The data are from E772 \protect\cite{E772}.  The dashed
line assumes a scaling linear in the atomic number A.  The shaded
region represents our calculations with a coherence time ranging
from 4 to 6 proton-nucleon collisions, both elastic and inelastic.
We computed with target nuclei C, Ca, Fe, W and Pb and interpolate
between with straight lines to guide the eye.  The value of Feynman
$x_F$ of the Drell-Yan pair is 0.65. }
\label{pady1}
\end{figure}

The experiment E772 \cite{E772} measured the ratio
$\sigma^{\rm DY}_{pA}/(\sigma^{\rm DY}_{pd}/2)$.  Were there no energy
loss and all nuclei were charge symmetric this ratio would be
equal to $A$.  The experiment measured muon pairs with invariant mass
$M$ between 4 and 9 GeV and greater than 11 GeV to eliminate
the $J/\psi$ and $\Upsilon$ contributions.  The data has been presented
in 7 bins of Feynman $x_F$ from 0.05 to 0.65.  (Recall that $x_F$ is
the ratio of the muon pair longitudinal momentum to the incident beam
momentum in the nucleon-nucleon c.m. frame.)  Data for an exemplary value
of $x_F$ is shown in Figure \ref{pady1}.
The data should fall on
the dashed line if the ratio of cross sections is $A$.  There is
a small but noticeable departure for tungsten and at the largest value
of $x_F$.  This is to be expected if energy loss plays a role as it
must affect the largest target nucleus and the highest energy muon
pairs the most \cite{enhance}.

We have computed the individual cross sections $\sigma^{\rm DY}_{pA}$
with a variable time delay.  The proton cascades through the nucleus
as described earlier, but we assume that the energy available to
produce a Drell-Yan pair is that which the proton has after $n$
previous collisions.  Thus $n = 0$ is full energy loss and $n = \infty$
is zero energy loss.  We have taken the resulting proton-nucleus
Drell-Yan cross section, multiplied it by 2, divided it by the
sum of the computed pp and pn cross sections and display the results
in Figure \ref{pady1}.  (For plots of other values of $x_F$ see \cite{prl}.)
The lower edge of the shaded regions in the figure
corresponds to $n = 4$ and the upper edge to $n = 6$.
Overall the best representation of the data lies in this
range.  This collision number shift is easily converted to a coherence
time.  Let $\tau_0$ be the coherence time in the c.m. frame of the
colliding nucleons.  This is essentially the same as the formation
time of a pion since most pions are produced with rapidities near
zero in that frame.  The first proton-nucleon collision is the most
important, so boosting this time into the rest frame of the target
nucleus and converting it to a path length (proton moves essentially
at the speed of light) gives $\gamma_{\rm cm}\,c\,\tau_0 \approx
\sqrt{\gamma_{\rm lab}/2}\,c\,\tau_0$.  This path length may then be
equated with $n$ times the mean free path
$l = 1/\sigma^{\rm tot}_{\rm NN} \rho$.  Using a total cross section
of 40 mb and a nuclear matter density of 0.155 nucleons/fm$^3$ we
obtain a path length of 8$\pm$2 fm and a proper coherence time
of 0.4$\pm$0.1 fm/c corresponding to $n=5\pm 1$.

This value of the proper coherence time is just about what should
have been expected {\it a priori}.  In the c.m. frame of the colliding
nucleons at the energies of interest a typical pion is produced
with an energy of E$_{\pi} \approx 500$ MeV.  By the uncertainty
principle this takes a time of order $\hbar c/E_{\pi} \approx 0.4$
fm/c.

\subsection{$J/\psi$ production and absorption}

A process related to Drell-Yan is the production of $J/\psi$ which we shall
address now.  This is also a relatively
hard process and so both energy loss of the beam proton and the
Landau-Pomeranchuk-Migdal effect must be taken into account. However, there
is an additional effect
which plays a role, and that is the occasional absorption or breakup of the
$J/\psi$ in encounters with target nucleons.  (The inelastic interaction of one
of the leptons in Drell-Yan production with target nucleons is ignorably
small.)  The absorption cross section, $\sigma_{\rm abs}$, has been estimated
in a straightforward Glauber analysis without energy loss and with an infinite
coherence/formation time to be about 6-7 mb \cite{DimaJ}.  This has formed the
basis for many analyses of $J/\psi$ suppression in heavy ion collisions.  Any
anomalous suppression may be an indication of the formation of quark-gluon
plasma \cite{matsui}, hence the importance of obtaining the most accurate
value of $\sigma_{\rm abs}$ possible.  This cross section has also been
inferred from photoproduction experiments of $J/\psi$ on nuclei from which a
value much less than that has been obtained \cite{photo}.  This has been a
puzzle. One attempt to resolve this apparent discrepancy consists of
modeling the produced $J/\psi$ state as a pre-resonant color dipole state
with two octet charges \cite{dima2}; however, the results are only
semi-quantitative.
\begin{figure}[htb!]
\centerline{\includegraphics[angle=90,width=4in]{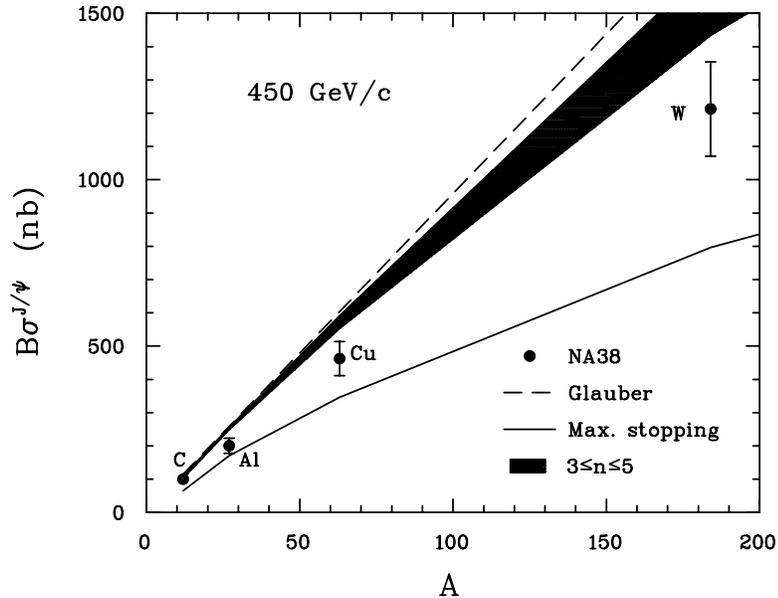}}
\caption{ Branching ratio into muons times cross section to produce $J/\psi$
with $x_F \ge 0$ in proton-nucleus collisions at 450 GeV/c.  The data is from
NA38 \protect\cite{Carlos}.  The dashed line is $A$ times the
nucleon-nucleon
production cross section.  The solid curve represents full energy
loss with zero coherence/formation time, while the banded region represents
partial energy loss
with a coherence/formation time within the limits set by Drell-Yan production.
(Computations were done for C, Al, Cu, W and U and the points connected by
straight lines to guide the eye.)}
\label{pajpsi1}
\end{figure}

In order to compute the
production cross section of $J/\psi$ in proton-nucleus collisions we need a
parametrization of it in the more elementary nucleon-nucleon collisions. For
this we call upon the parametrization of a compilation of data by
Louren\c{c}o \cite{Carlos}.
\begin{equation}
B \sigma_{NN \rightarrow J/\psi}(x_F > 0) = 37 \left( 1 - m_{J/\psi}/\sqrt{s}
\right)^{12} \, {\rm nb}
\end{equation}
Here $B$ is the branching ratio into dimuons and $x_F$ is the ratio of the
momentum carried by $J/\psi$ to the beam momentum in the center of mass frame
$(-1 < x_F < 1)$.  Due to the degradation in momentum of the proton as it
traverses the nucleus it is important to know the $x_F$ dependence of the
production.  The Fermilab experiment E789 has measured this dependence at 800
GeV/c \cite{E789} to be proportional to $(1-|x_F|)^5$.  Assuming that this
holds at lower energy too we use the joint $\sqrt{s}$ and $x_F$
functional dependence and magnitude:
\begin{equation}
\frac{d\sigma_{NN \rightarrow J/\psi}}{dx_F} = 6 \sigma_{NN \rightarrow
J/\psi}(x_F > 0) (1-|x_F|)^5 \, .
\end{equation}
The cross section in proton-nucleus collisions can now be computed in LEXUS
with no ambiguity.

Figure \ref{pajpsi1} shows the results of our calculation in 
comparison to data taken by NA38 \cite{Carlos}.
The dashed curve is $A$ times the nucleon-nucleon production cross section;
it obviously overestimates the data.  The
solid curve shows the result of LEXUS with full energy degradation of the beam
proton without account taken of the Landau-Pomeranchuk-Migdal effect; it 
obviously underestimates the data.  The hatched region represents the 
inclusion
of the latter effect with a proper formation/coherence time $\tau_0$ in the
range of 0.3 to 0.5 fm/c consistent with Drell-Yan production.  The
time delay is implemented as described previously:  the energy 
available for the
production of $J/\psi$ is that which the proton had $n$ collisions prior; that
is, the previous $n$ collisions are ignored for the purpose of determining the
proton's energy. This is an approximate treatment of the
Landau-Pomeranchuk-Migdal effect.  The $n$ is related to the beam energy and
to  the coherence time $\tau_0$ in the center of mass frame of the
colliding nucleons.  Using, as before, a total cross section
of 40 mb, a nuclear matter density of 0.155 nucleons/fm$^3$, and
$0.3 < \tau_0 <
0.5$ fm/c we obtain $2 < n < 3$ at 200 GeV/c and $3 < n < 5$ at 450 GeV/c. As
may be seen from the figure, the data is overestimated, indicating the
necessity for nuclear absorption.

\begin{figure}[htb!]
\centerline{\includegraphics[angle=90,width=4in]{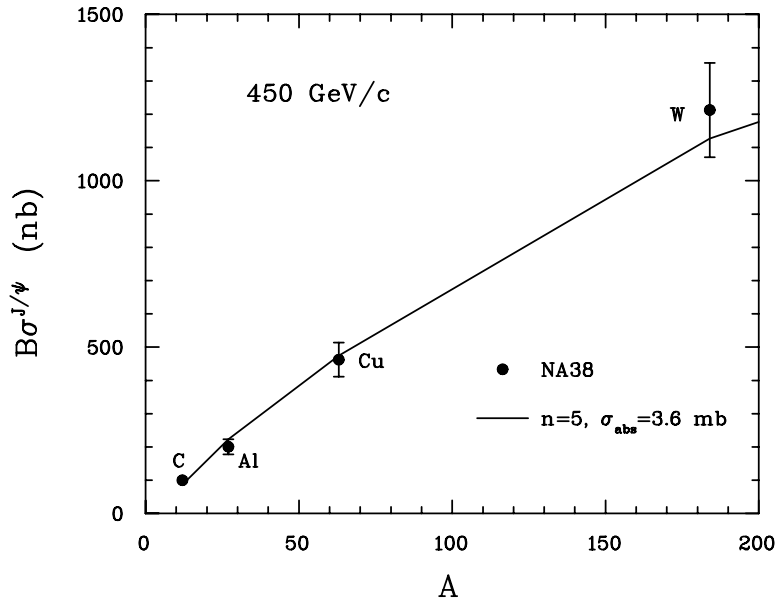}}
\caption{ Same data as in Figure \ref{pajpsi1}.  The solid
curve is the best fit of the
model which includes beam energy loss with a coherence time of 0.5 fm/c (n=5 at
this energy) and a $J/\psi$ absorption cross section of 3.6 mb. }
\label{bestfit}
\end{figure}

We now introduce a $J/\psi$ absorption cross section on nucleons and compute
its effect within LEXUS in the canonical way \cite{DimaJ,plb}.  When 
the $J/\psi$
is created there will in general be a nonzero number of nucleons blocking its
exit from the nucleus.  Knowing where the $J/\psi$ is created allows one to
calculate how many nucleons lie in its path, and hence, to compute the
probability that it will be dissociated into open charm.  We choose a value
of $\tau_0$ allowed by Drell-Yan measurements, mentioned above, and then
vary $\sigma_{\rm abs}$, assuming that it is energy
independent.  The lowest value of chi-squared for the 200 and 450 GeV/c
data set taken together is obtained with $\tau_0 = 0.5$ fm/c and
$\sigma_{\rm abs} = 3.6$ mb.  The results are shown in Figure \ref{bestfit}.
The fitted values all lie within one standard deviation of the data
points.  This is quite a satisfactory representation of the data.  It means
that both Drell-Yan and $J/\psi$ production in high energy proton-nucleus
collisions can be understood in terms of a conventional hadronic
analysis when account is taken of the energy loss of the beam proton, the
Landau-Pomeranchuk-Migdal effect, and nuclear absorption of the $J/\psi$
in the final state.  It also means that the absorption cross section for
$J/\psi$ inferred from high energy proton-nucleus collisions is
consistent with the value inferred from photoproduction experiments on nuclei.

\section{NUCLEUS-NUCLEUS COLLISIONS}

We are now ready to apply our model to nucleus-nucleus
collisions. Recall that the coherence time is determined by comparing to high
energy proton-nucleus Drell-Yan data, and that the $J/\psi$ nuclear
absorption is determined by comparing to proton-nucleus results. Those
parameters are now kept fixed.

\begin{table}[htb]
\caption{\ }
\label{table1}
\newcommand{\m}{\hphantom{$-$}}
\newcommand{\cc}[1]{\multicolumn{1}{c}{#1}}
\renewcommand{\tabcolsep}{1pc} 
\renewcommand{\arraystretch}{1.2} 
\begin{tabular}{@{}llllll}
\hline
\ & Process & $\sigma_{\rm expt.}^{\rm tot}$ & \cc{n=$\infty$} & \cc{n=3} 
& \cc{n=2} \\
\hline
S+U: &  & & & &  \\
\ & Drell-Yan & 310 $\pm$ 10 $\pm$ 25 nb  & \m443 & \m323 & \m267      \\
\ & $J/\psi$ & 7.78 $\pm$ 0.04 $\pm$ 0.62 $\mu$b  & \m11.5 & \m7.92  
& \m6.46 \\
Pb+Pb: & & & & & \\
\ & Drell-Yan & 1.49$\pm$ 0.01 $\pm$ 0.11 $\mu$b  & \m2.13 & \m1.32 & \m1.02 \\
\ & $J/\psi$ & 21.9$\pm$ 0.02 $\pm$ 1.6 $\mu$b  & \m45.5 & \m26.7 & \m20.2 \\
\hline
\end{tabular}\\[2pt]
The S+U processes are measured by NA38 \cite{NA38totsig} in collisions 
at 200 GeV/c, and
the Pb+Pb processes are measured by NA50 \cite{NA50totsig} in
collisions at 158 GeV/c. 
\end{table}

\begin{figure}[htb!]
\centerline{\includegraphics[width=3in]{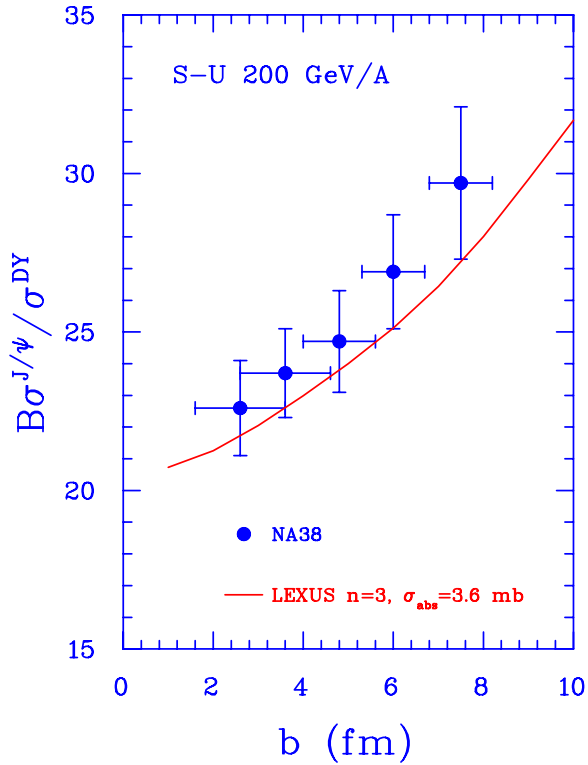}}
\caption{ Ratio of $J/\psi$ to Drell-Yan production cross sections. $B$
is the branching ratio into a muon pair. The data is from 
Reference \cite{NA38totsig}.}
\label{SUratio}
\end{figure}
In the kinematical region appropriate for the CERN nucleus-nucleus
experiments we find the absolute total cross-sections given in Table
\ref{table1}.
At the energies relevant for the table, 
the kinematical arguments presented in section \ref{sect:pA} tell us
that we should have $2\le~n\le~3$. 
From the table, we see
that the calculations satisfy this condition for the data involving the
S projectile. With the heavier system, the measured $J/\psi$ 
production cross section just lies in the required interval, while the
Drell-Yan value is slightly underestimated. Notice in particular that 
the calculations
with no energy loss ($n = \infty$)  overshoot all the 
experimental values.
\begin{figure}[htb!]
\centerline{\includegraphics[width=3in]{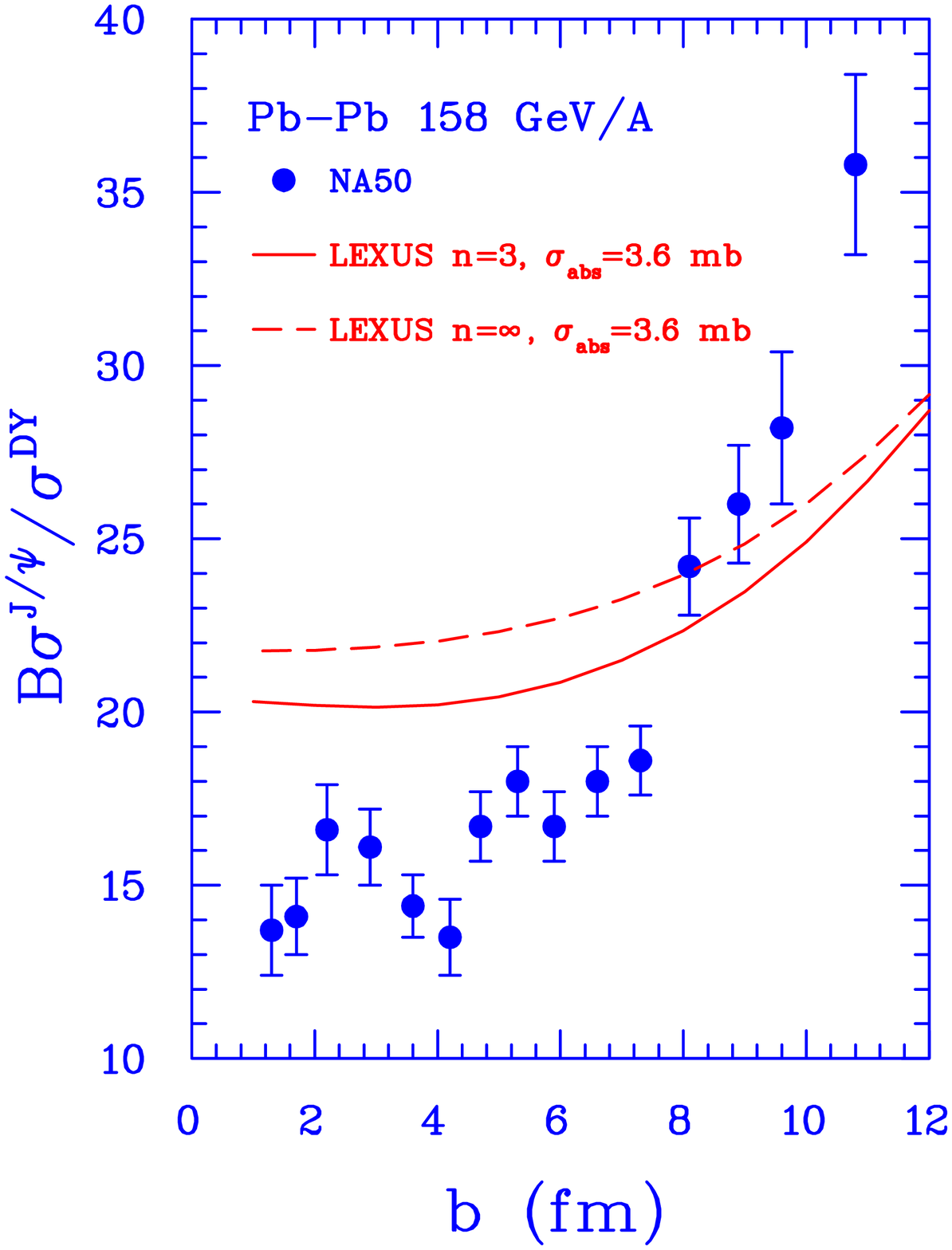}}
\caption{ Ratio of $J/\psi$ to Drell-Yan production cross sections. $B$
is the branching ratio into a muon pair. The data is from 
Reference \cite{NA50thresh}.}
\label{Pbratio}
\end{figure}

We can also calculate the $J/\psi$/Drell-Yan ratio and compare the 
results to NA38 and NA50 measurements. Turning first to S+U
collisions at 200 GeV/c, we plot the results of the calculation
together with the experimental data in Figure \ref{SUratio}. We use the
value of $n$ suggested by minimizing $\chi^2$ in an analysis of
proton-nucleon data; see Figure \ref{bestfit} and Reference \cite{plb}. 
The impact parameter range corresponding to a given bin in
transverse energy is extracted by the experimental collaboration
\cite{NA38totsig}. One sees that the measurements are reproduced by the 
model without any new parameters.  There is some room for improvement
within the experimental constraints we have chosen for ourselves;
however, such fine tuning is not warranted here and we will postpone this
exercise \cite{gjkprep}.
Applying the model to the Pb+Pb data from NA50 \cite{NA50thresh}, we
get the full curve shown in Figure \ref{Pbratio}. 
Clearly, these experimental data are not reproduced in trend nor in
magnitude. Bear in mind, however, that the quality of this fit is very
comparable to other calculations with purely hadronic scenarios
\cite{kluberg}. In Figure \ref{Pbratio}, the dashed line shows the same
ratio without energy loss. The impact parameter has been
extracted by the experimental collaboration. 

\section{CONCLUSION}

The analysis performed here can and should be improved upon.  What
we have done is a rough approximation to adding the quantum mechanical
amplitudes for a proton scattering from individual nucleons within
a nucleus.  A more sophisticated treatment would undoubtedly lead
to even better agreement with experiment, but the inferred value of
the proper coherence time is unlikely to be much different than obtained
with this first estimate.  It will be very instructive to repeat this
analysis in the language of partonic variables.  Actually, the analysis
with parton energy loss alone was reported by Gavin and Milana \cite{sean}
with satisfactory results obtained for Drell-Yan if the quarks/antiquarks
lose about 1.5 GeV/fm.  Nuclear shadowing \cite{shadows} needs to be taken
into account too.  The
relationship among all these effects is not well-understood, nor is the
relationship between these effects in partonic and hadronic variables.

Our calculation of the $J/\psi$-to-Drell-Yan ratio is in agreement
with the data for the light system, but fails with the Pb+Pb system. 
A systematic exploration of the freedom allowed by present
parametrizations of the nucleon-nucleon experimental data is called for
\cite{gjkprep}.  It may be that after a more thorough treatment, the need for
a picture with $J/\psi$ absorption on ``co-movers'' \cite{DimaJ,comovers} 
will emerge here also. This, 
and the issues enumerated above are being investigated.

\section{ACKNOWLEDGEMENTS}

This work was supported by the U. S. Department of Energy under grant
DE-FG02-87ER40328, by the Natural
Sciences and Engineering Research Council of Canada, by the Fonds FCAR of
the Quebec Government, by the Director, 
Office of Energy Research, Office of High Energy and Nuclear Physics, 
Division of Nuclear Physics, and by the Office of Basic Energy
Sciences, Division of Nuclear Sciences, of the U.S. Department of
Energy 
under contract No. DE-AC03-76SF00098.


\begin{thebibliography} {20}
\bibitem{enhance} There is actually an enhancement of this ratio
near $x_F = 0$, and possibly for negative $x_F$ although it hasn't
been measured by E772 in the backward direction.  Such an
enhancement may be expected due to Drell-Yan production in occasional
collisions of secondary particles with target nucleons.
\end{thebibliography}
\end{document}